\documentclass[12pt,preprint]{aastex}

\slugcomment{ver. 2.1}

\shorttitle{A Structured Leptonic Jet Model}
\shortauthors{Kusunose, \& Takahara}

\begin{document}
\newcommand{\D}{\displaystyle}

\title{A Structured Leptonic Jet Model 
of the ``Orphan'' TeV Gamma-Ray Flares in TeV Blazars}

\author{Masaaki Kusunose}
\email{kusunose@kwansei.ac.jp}
\affil{Department of Physics, School of Science and Technology,
Kwansei Gakuin University, Sanda 669-1337, Japan}

\and

\author{Fumio Takahara}
\affil{Department of Earth and Space Science,
Graduate School of Science, Osaka University,
Toyonaka 560-0043, Japan}
\email{takahara@vega.ess.esi.osaka-u.ac.jp}

\begin{abstract}
The emission spectra of TeV blazars extend up to tens of TeV and
the emission mechanism of the TeV $\gamma$-rays is explained by
synchrotron self-Compton scattering in leptonic models.
In these models the time variabilities of X-rays and TeV $\gamma$-rays
are correlated.
However, recent observations of 1ES 1959+650 and Mrk 421 have found 
the ``orphan'' TeV $\gamma$-ray flares, i.e.,
TeV $\gamma$-ray flares without simultaneous X-ray flares.
In this paper we propose a model for the ``orphan'' TeV $\gamma$-ray flares,
employing an inhomogeneous leptonic jet model.
After a primary flare that accompanies flare-up both in X-rays and
TeV $\gamma$-rays, radiation propagates in various directions in
the comoving frame of the jet.
When a dense region in the jet receives the radiation, X-rays
are scattered by relativistic electrons/positrons to become TeV $\gamma$-rays.
These $\gamma$-ray photons are observed as an ``orphan'' TeV 
$\gamma$-ray flare.  The observed delay time between 
the primary and ``orphan'' flares is about two weeks 
and this is accounted for in our model for parameters such as
$\Gamma = 20$, $d = 4 \times 10^{17}$cm, $\alpha = 3$, and $\eta = 1$,
where $\Gamma$ is the bulk Lorentz factor of the jet, 
$d$ is the distance between the central black hole and the primary flare site, 
$\alpha/\Gamma$ is the angle between the jet axis 
and the direction of the motion of the dense region 
that scatters incoming X-rays produced by the primary flare, 
and $\eta/\Gamma$ is the angle between the jet axis and the line of sight.
\end{abstract}


\section{Introduction}

Blazars are a subclass of active galactic nuclei and their high energy emission
is generated in regions near the central super massive black holes.
Because of their intense and rapidly variable radiation,
the emission regions are thought to be in relativistic jets 
closely aligned to the line of sight \citep[see][for a review]{aha04,kra04}.
The spectral energy distributions (SEDs) of blazars are characterized by
double broad peaks in the $\nu$-$\nu F_\nu$ representation.
One peak is in the optical to X-ray regions and the other is in
the $\gamma$-ray region.
Very high energy (100 MeV -- 1 GeV) $\gamma$-rays from blazars were
discovered by {\it EGRET} (Energetic Gamma-Ray Experiment Telescope) on board
Compton Gamma-Ray Observatory \citep{har99}.
Recent observations by Cerenkov telescopes have revealed that
many blazars emit $\gamma$-rays up to tens of TeV
\citep[][for a review]{aha05hess-all}.
Flare activities in X-rays and $\gamma$-rays from blazars are also known.
Although the emission mechanisms of blazars are still under study,
the SEDs of blazars are well explained by leptonic or hadronic models.
In the framework of the leptonic models,
the emission in the optical to X-ray region is attributed to
synchrotron radiation by nonthermal electrons/positrons in the jet
and the TeV $\gamma$-rays are accounted for
by the synchrotron self-Compton (SSC) model \citep{mgc92}.
The SSC model assumes that synchrotron photons are inverse-Compton 
scattered by the same nonthermal electrons/positrons 
that emit synchrotron photons.
If the SSC model is applied, the time variabilities of the X-rays 
and $\gamma$-rays should be correlated \citep[e.g.,][]{mkir97,lk00,kus00},
and indeed most flares occur almost simultaneously in X-rays and $\gamma$-rays.
However, the recent observations of 1ES 1959+650 \citep{kra-et04,dan05}
found that there is a TeV $\gamma$-ray flare
that is not accompanied by a X-ray flare,
which was observed on June 4, 2002 (MJD 52,429) by the Whipple telescope.
This is called the ``orphan'' TeV $\gamma$-ray (OTG) flare.
The OTG flare occurred about 15 days after a usual flare
in which X-ray and $\gamma$-ray flares were observed contemporaneously.
More recently another OTG flare might have been
observed from Mrk 421 \citep{bla05},
although the X-ray flux seems to have peaked about 1.5 days
before this TeV flare and this may not be a true OTG flare.

The existence of the OTG flares is challenging for the leptonic models.
\citet{bot05} has recently proposed a model for the OTG flare
of ES 1959+650, i.e., the hadronic synchrotron mirror model.
He assumed that a fraction of X-rays produced in the primary flare
by synchrotron radiation of leptons is reflected 
by a plasma cloud with scattering depth $\sim 0.1$
located in the direction of the jet propagation.
The reflected X-rays collide with protons in the jet 
and pions are produced.
Subsequently a large number of neutral pions decay into $\gamma$-rays, 
which are expected to be observed as TeV $\gamma$-rays.
If this model is applied to 1ES 1959+650, it is found that
the reflecting plasma is located at a distance 
$\sim 3 (\Gamma/10)^2 \Delta t_{20}$ pc from the central black hole, 
where $\Gamma$ is the bulk Lorentz factor of the jet and 
$\Delta t_{20} = \Delta t/(20 \, \mathrm{days})$ is the normalized 
time delay between the primary and OTG flares.
This model requires unreasonably large values of proton density in the jet
and the hadronic jet power (B\"{o}ttcher 2006: erratum).
Thus the hadronic synchrotron mirror model may not be applicable to OTG
flares,
although it may be still viable if the effects that reflected 
synchrotron photons increase as the blob approaches
to the mirror are taken into account.
(More recently \citet{rbp05} calculated the predicted neutrino spectrum due to
the decay of charged pions in the framework of the hadronic synchrotron mirror
model.)

In this paper, we propose another model of the OTG flares in the framework
of the SSC model, assuming that the emission region is not homogeneous,
which is different from the conventional SSC model.
We assume that the injection of nonthermal electrons/positrons
triggers the primary flare where both X-rays and TeV $\gamma$-rays
are emitted.
If the jet is not uniform but consists of a few patchy regions,
X-rays that were produced in the primary flare impinge on 
another dense region of the jet.
This sudden increase of X-ray photons results in strong 
TeV $\gamma$-ray flux by inverse Compton scattering,
which is expected to be observed as an OTG flare.
Because there is a time delay between the primary flare in one region
and the scattering in another region, 
TeV $\gamma$-rays are observed as an OTG flare.
Note that TeV $\gamma$-rays from the primary flare are not
scattered in the OTG flare site because of Klein-Nishina effects.
The observed delay time between the OTG flare and 
the previous flare is about two weeks in 1ES 1959+650 \citep{kra-et04}.
The delay time between the primary and OTG flares 
in the comoving frame of the jet is much shorter 
than the synchrotron cooling time 
of nonthermal electrons/positrons (see \S \ref{sec:lumi} for detail),
if the magnetic field in the jet is about 0.1 G.
Although high energy electrons/positrons injected into the primary flare
are rapidly cooled,
nonthermal electrons/positrons that emit TeV $\gamma$-rays 
in a quiescent state may be still continuously injected into the jet
and they may contribute to the secondary flare.
In this paper we do not specify the acceleration mechanisms of those 
electrons/positrons or calculate the emission spectrum of the flare
but focus our work on the kinematics of the jet.

The kinematics of the jet is described in \S \ref{sec:kinematics} and 
the jet properties are presented in \S \ref{sec:lumi}.
A summary of our results and discussion are given in \S \ref{sec:sum}.


\section{Kinematics of Jets} \label{sec:kinematics}

We first study the kinematics of a jet shown in Figure \ref{fig:jet-case1}
(Case 1).
We assume that a jet (a patchy plasma shell) with bulk Lorentz factor 
$\Gamma$ is ejected radially from point O that is close to 
the central black hole.
The observer is at angle $\theta_\mathrm{obs} = \eta / \Gamma$ 
from the jet axis, where $\eta \sim {\cal O}(1)$.
At P the primary flare occurs, which is triggered by, e.g., the internal shock.
Here we assume that the high energy particles are confined in a patchy shell
and high density regions are located at P and S$_\mathrm{P}$ in the shell.
The emission mechanisms in a flare are synchrotron emission and SSC
for X-rays and TeV $\gamma$-rays, respectively.
A fraction of the emitted photons propagate along directions that are 
slightly off from the jet axis and they are injected into the high density
region around S, where S was located at S$_\mathrm{P}$ when the primary
flare occurred.
The injected TeV $\gamma$-rays are in Klein-Nishina regime and are
not scattered but X-rays are Compton-scattered to become TeV $\gamma$-rays.
This triggers the secondary (OTG) flare in the TeV region.

\begin{figure}
\includegraphics[angle=0,scale=.65]{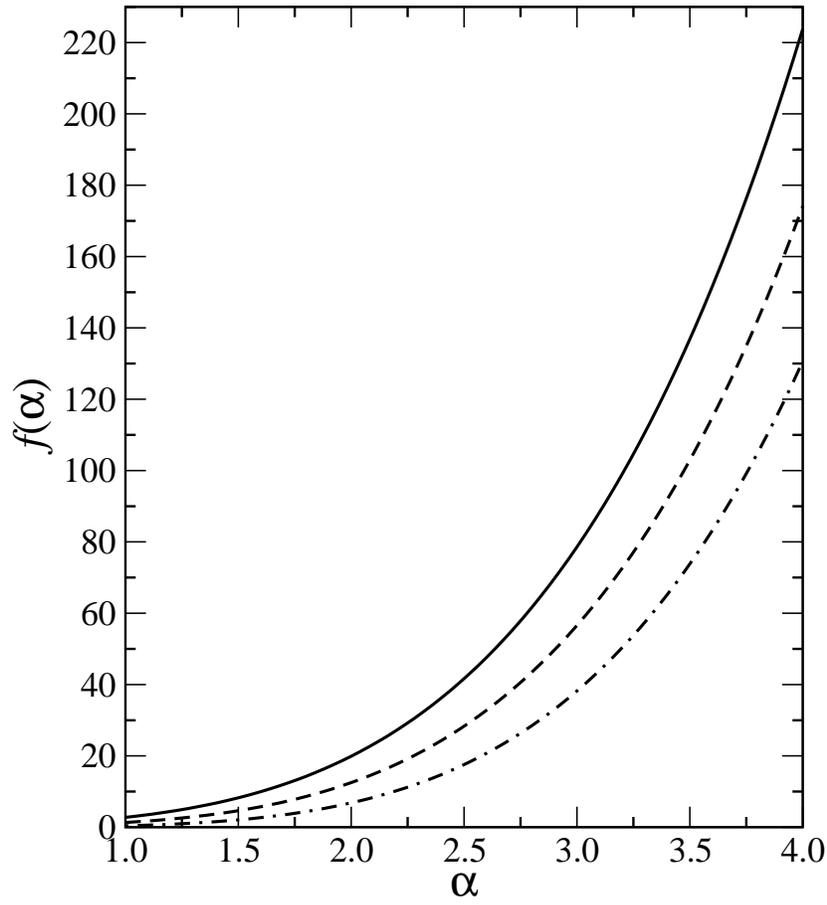}
\caption{$f(\alpha)$ for $\eta = 0$ (solid), 0.5 (dashed), 
and 1 (dot-dashed).
\label{fig:f-factor}}
\end{figure}

We assume that the photons emitted at P reach S after time interval
$\delta t$ in the rest frame of the black hole
and that the propagation direction of those photons is denoted by $\theta_f$.
For the geometry shown in Figure \ref{fig:jet-case1},
equations to be solved for $\delta t$ and $\theta_f$ for given
$d$, $\xi$, and $\Gamma$ are as follows:
\begin{equation}
\label{eq:sin}
(d + \beta \, c \, \delta t) \sin \xi
= c \, \delta t \,\sin \theta_f ,
\end{equation}
and
\begin{equation}
\label{eq:cos}
(d + \beta \, c \, \delta t)^2 = d^2 + (c \,\delta t)^2 
- 2 \, d \, c \, \delta t \, \cos (\pi - \theta_f) ,
\end{equation}
where $c$ is the light speed and $\beta = \sqrt{1 - 1/\Gamma^2}$.
Here we assumed that the distances between O and P, OP, 
and O and S$_\mathrm{P}$, OS$_\mathrm{P}$, are the same, i.e.,  
$\mathrm{OP} = \mathrm{OS}_\mathrm{P} = d$.
Next we assume that 
$\xi = \alpha/\Gamma$, where $\alpha \sim {\cal O}(1)$ is a constant.
Using equations (\ref{eq:sin}) and (\ref{eq:cos}), we obtain
the equation for $x \equiv c \delta t/d$:
\begin{equation}
x^2 + 4 \Gamma \sqrt{\Gamma^2-1} \left( \frac{2 \alpha^2}{x} - x \right)
+ \frac{4 \alpha^2 \Gamma^2}{x^2} - 4 \alpha^2 + 4 (\alpha^2-1) \Gamma^2 = 0 .
\end{equation}
There are four solutions for $x$, 
but the physically allowed solution is unique and given by
\begin{eqnarray}
x &=& \Gamma \sqrt{\Gamma^2-1} 
- \sqrt{\Gamma^4-(1+\alpha^2)\Gamma^2+\alpha^2} \nonumber \\
& &+\left\{ 2 \Gamma^4 -\alpha^2 (\Gamma^2-1)
+ \frac{2 \Gamma^3  (\Gamma^2-1)^{1/2}}
{[\Gamma^4-(1+\alpha^2)\Gamma^2+\alpha^2]^{1/2}}
(\alpha^2 - \Gamma^2) \right\}^{1/2} .
\label{eq:sol-x}
\end{eqnarray}
When $\Gamma \gg 1$, $c \, \delta t/d$ is approximated as
\begin{equation}
\label{eq:sol-app}
\frac{c \, \delta t}{d} \approx \frac{\alpha^2}{2}
+ \sqrt{\alpha^2 + \frac{\alpha^4}{4}} + {\cal O} (\Gamma^{-2}) .
\end{equation}
From equation (\ref{eq:sin}) we obtain
\begin{equation}
\sin \theta_f \approx \frac{1}{\Gamma} \left[ \alpha 
+ \frac{2}{\alpha + \sqrt{4+\alpha^2}} \right] .
\label{eq:sol-sin}
\end{equation}
Because the angle between the jet axis and the line of sight is
$\theta_\mathrm{obs}$,
the observed delay time of the secondary flare from the primary flare
is given by
\begin{equation}
\delta t_\mathrm{obs} = [1 - \cos ( \theta_f - \theta_\mathrm{obs} )]
\, \delta t 
\approx \frac{1}{\Gamma^2} \, f(\alpha) \, \frac{d}{c} \, ,
\end{equation}
where
\begin{equation}
f(\alpha) = \frac{1}{4} \left(\alpha^2+\sqrt{\alpha^2+\frac{\alpha^4}{4}}
\right) \left[ 2 + \alpha^2 + \sqrt{4 \alpha^2 + \alpha^4}
- \alpha \eta \left( 4 + \frac{8}{\alpha^2 + \sqrt{4 \alpha^2
+ \alpha^4}} \right) + 2 \eta^2 \right] .
\end{equation}
The behavior of $f(\alpha)$ is shown in Figure \ref{fig:f-factor}.
For example, $f \approx 38.2$ for $\alpha = 3$ and $\eta = 1$.
Note the strong dependence of $f$ on $\alpha$.
The numerical value of $\delta t_\mathrm{obs}$ is estimated as
\begin{equation}
\delta t_\mathrm{obs} \approx 
0.29 \left( \frac{20}{\Gamma} \right)^2
\left( \frac{d}{3 \times 10^{17} \, \mathrm{cm} } \right) f(\alpha) 
\quad \mathrm{days} .
\end{equation}
Then $\delta t_\mathrm{obs} \approx 15$ days is obtained for $\Gamma = 20$,
$d = 4 \times 10^{17}$ cm, $\eta = 1$, and $\alpha = 3$.
This value is close to the observed delay time of the OTG flare
in 1ES 1959+650 \citep{kra-et04}.
These parameters give the distance of S from the central black hole
$d + \beta c \delta t \approx 1.4$ pc.

\begin{figure}
\includegraphics[angle=0,scale=.65]{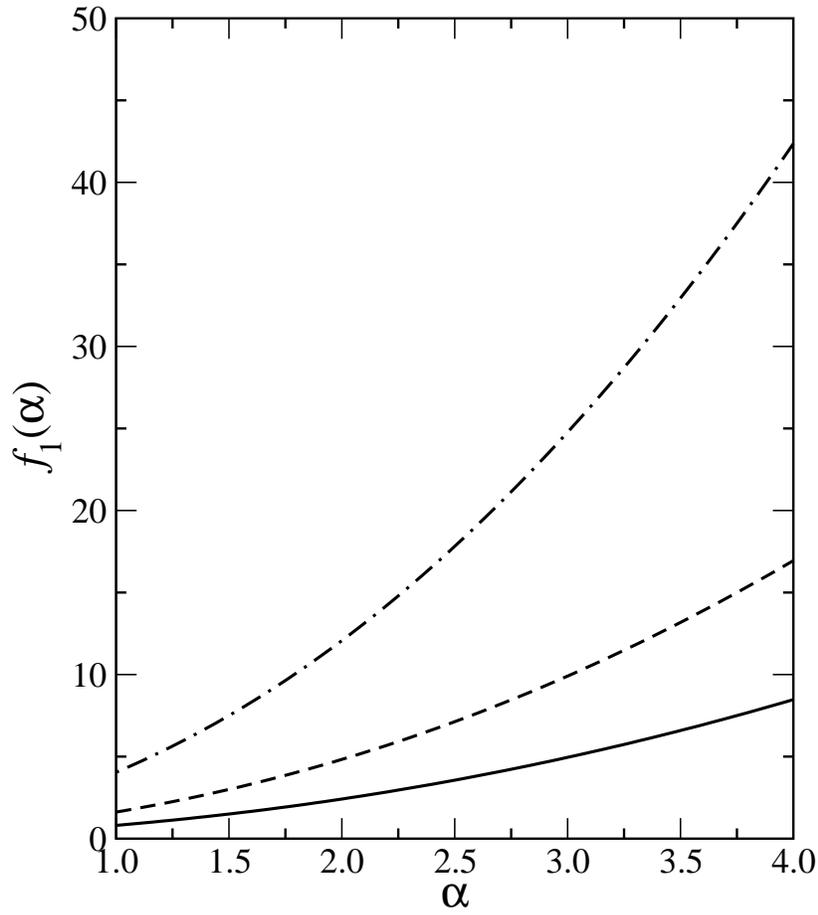}
\caption{$f_1(\alpha)$ for $\eta = 0$ (solid), 1 (dashed),
and 2 (dot-dashed).
\label{fig:f1-geom2}}
\end{figure}

There is another possibility of the jet geometry (Case 2) as shown in 
Figure \ref{fig:jet-case2}.
Here we assume that the X-ray source is located
at P that is away from the jet axis and that the secondary flare occurs
at S$_\mathrm{P}$ on the jet axis.
The solution for $c \delta t/d$ is the same as equation (\ref{eq:sol-app})
and $\theta_f$ is given by equation (\ref{eq:sol-sin}).
If a flare at S$_\mathrm{P}$ occurs simultaneously with a flare at P
and the flare at S$_\mathrm{P}$ is observed,
the observed time delay between the flares at S$_\mathrm{P}$ and S is
given by 
\begin{equation}
\label{eq:tout1}
\delta t_\mathrm{obs} = (1 - \beta \cos \theta_\mathrm{obs}) \delta t
\approx \frac{1}{\Gamma^2} \, f_1(\alpha) \, \frac{d}{c} ,
\end{equation}
where 
\begin{equation}
f_1(\alpha) = \left( \frac{\alpha^2}{2} 
+ \sqrt{\alpha^2 + \frac{\alpha^4}{4}}\right) \frac{1 + \eta^2}{2} .
\end{equation}
The values of $f_1(\alpha)$ for $\eta = 0$, 1, and 2 are
shown in Figure \ref{fig:f1-geom2}.
If $\Gamma = 20$, $\alpha = 3$, $\eta = 1$,
and $d = 4 \times 10^{17}$ cm are assumed,
the delay time is about 4 days and too short to explain the observed
delay time.  If $\eta$ is larger, the delay time becomes larger,
but the observed luminosity decreases significantly
because of the beaming effects.

\begin{figure}
\includegraphics[angle=0,scale=.65]{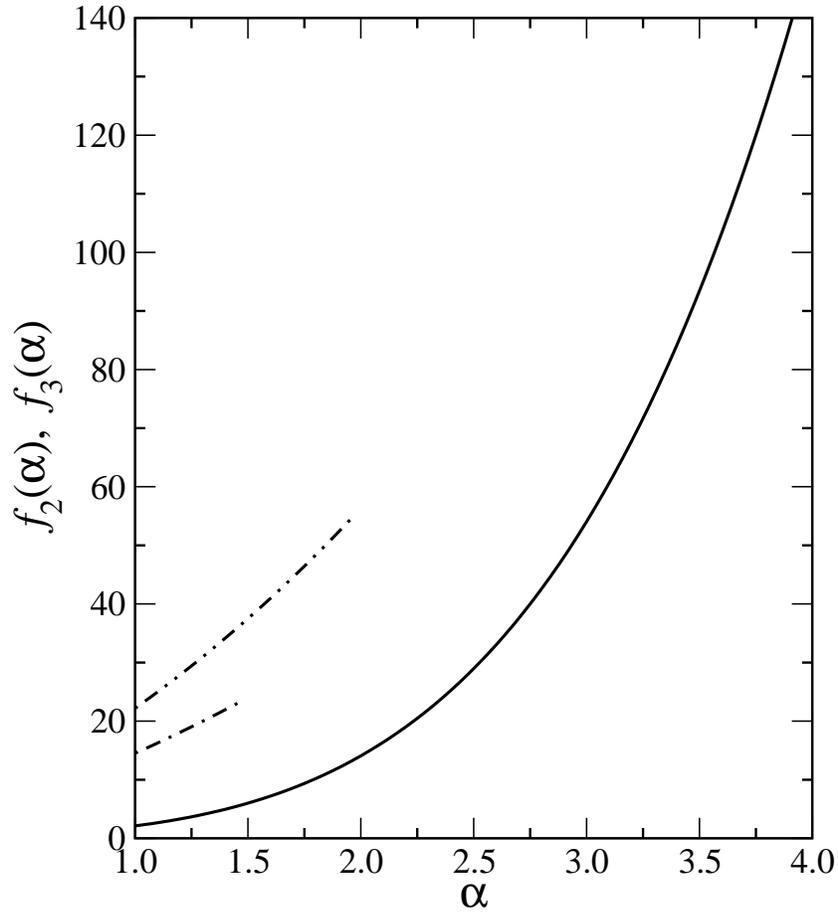}
\caption{$f_2(\alpha)$ for $\eta = 1.5$ (dot-dashed) 
and 2 (dot-dot-dashed) and
$f_3(\alpha)$ (solid). $f_2(\alpha)$ and $f_3(\alpha)$ are
used to calculate $\delta t_\mathrm{obs}$ in Eq. (\ref{eq:tobs-case2})
for $\alpha < \eta$ and $\alpha > \eta$, respectively.
\label{fig:f2-3-geom2}}
\end{figure}

If the primary flare at P is observed and a flare at S$_\mathrm{P}$ is
not observed, the delay time between the flares at P and S is given by
\begin{equation}
\label{eq:tobs-case2}
\delta t_\mathrm{obs} =
\cases{
[1-\cos(\theta_\mathrm{obs} - \xi + \theta_f)] \delta t 
\approx \frac{\D 1}{\D \Gamma^2} \, f_2(\alpha) \, \frac{\D d}{\D c}
& if $\xi < \theta_\mathrm{obs}$ \cr
(1-\cos \theta_f) \delta t 
\approx \frac{\D 1}{\D \Gamma^2} \,f_3(\alpha) \, \frac{\D d}{\D c}
& if $\xi > \theta_\mathrm{obs}$ ,
}
\end{equation}
where
\begin{equation}
f_2(\alpha) = 4 \alpha \eta
+ \alpha^2 (\eta^2-1) + \alpha \sqrt{4 + \alpha^2}
(1 + \eta^2) ,
\end{equation}
and \begin{equation}
f_3(\alpha) =  \frac{\alpha^4 + 3 \alpha^2}{4}
+ \frac{1 + \alpha^2}{2} \sqrt{\alpha^2 + \frac{\alpha^4}{4}} .
\end{equation}
$f_2(\alpha)$ for $\eta =1.5$ and 2
and $f_3(\alpha)$ are shown in Figure \ref{fig:f2-3-geom2}.
When $\xi < \theta_\mathrm{obs}$,
$\eta \gtrsim 1.5$ and $\alpha \gtrsim 1.5$ are needed 
to obtain $\delta t_\mathrm{obs} \sim$ 2 weeks,
if $\Gamma = 20$ and $d = 4 \times 10^{17}$ cm are assumed.
When $\xi > \theta_\mathrm{obs}$,
$\alpha \gtrsim 2.7$ is required to explain $\delta t_\mathrm{obs} \sim$ 
2 weeks for the same parameter values as the above.

\section{Properties of the Secondary Flare} \label{sec:lumi}

\subsection{Timescales}

The solution obtained in \S \ref{sec:kinematics}
gives the distance, S$_\mathrm{P}$S, that the emission region travels during
$\delta t$ as
\begin{equation}
\beta c \delta t \sim
\left( \frac{\alpha^2}{2} + \sqrt{\alpha^2 + \frac{\alpha^4}{4}} \right) d .
\end{equation}
In the comoving frame of the jet, the travel time is given by
\begin{equation}
\delta t' \sim \frac{\delta t}{\Gamma}
\approx 57.9
\left( \frac{\alpha^2}{2} + \sqrt{\alpha^2 + \frac{\alpha^4}{4}} \right) 
\left(\frac{20}{\Gamma}\right) 
\left( \frac{d}{3 \times 10^{17} \mathrm{cm}} \right)
\quad \mathrm{days} .
\end{equation}
If $\alpha = 3$, we obtain $\delta t' \sim 574$ days.
Here primed quantities are evaluated in the comoving frame of the jet.

Synchrotron cooling rate per electron in the comoving frame of the jet
is given by
$P'_\mathrm{syn} = (4/3) \sigma_\mathrm{T} c \beta^{\prime \, 2}
\gamma^{\prime \, 2} u'_B$, 
where $u'_B$ is the magnetic energy density.
Then we obtain the cooling time as
$t'_\mathrm{syn} = \gamma' m_e c^2/P'_\mathrm{syn}$.
The characteristic energy of synchrotron radiation is given by
$\epsilon'_\mathrm{syn} = 3 \hbar e B' \gamma^{\prime \, 2} /(2 m_e c)$
and the observed synchrotron photon energy is Doppler shifted, i.e.,
$\epsilon_\mathrm{syn} \sim \Gamma \epsilon'_\mathrm{syn}$.
Because synchrotron photons with energy $\epsilon'_\mathrm{syn}$ 
are emitted by electrons with a Lorentz factor
\begin{equation}
\label{eq:gamma-value}
\gamma' \sim 5.37\times 10^5
\left( \frac{B'}{0.1 \mathrm{G}} \right)^{-1/2} \, 
\left( \frac{\Gamma}{20} \right)^{-1/2} \, 
\left( \frac{\epsilon_\mathrm{syn}}{10 \mathrm{keV}} \right)^{1/2} ,
\end{equation}
we obtain
\begin{equation}
t'_\mathrm{syn} 
\sim 1.7 \left(\frac{B'}{0.1 \mathrm{G}} \right)^{-3/2}
\left(\frac{\Gamma}{20} \right)^{1/2}
\left( \frac{\epsilon_\mathrm{syn}}{10 \mathrm{keV}} \right)^{-1/2} 
\quad \mathrm{days} .
\end{equation}
Then we find that $t'_\mathrm{syn} < \delta t'$.
Because the cooling by inverse Compton scattering is not negligible,
the cooling time is much shorter than $t'_\mathrm{syn}$.
Thus relativistic electrons/positrons injected 
into the primary flare are already cooled when the secondary flare occurs.
For the secondary flare to be caused, the injection of
relativistic electrons/positrons is necessary.
The injection rate, however, does not need to 
be as high as in the primary flare.
When synchrotron photons from the primary flare impinge on the dense region, 
which is located at the right place as described in \S \ref{sec:kinematics},
the secondary flare occurs by inverse Compton scattering of those photons.


\subsection{Beaming effects}

There is a slight velocity difference between S$_\mathrm{P}$ and P.
The relative velocity between points S$_\mathrm{P}$ and P normalized 
by the light speed is given by
$\beta_\mathrm{rel} = 2 \beta \sin (\xi/2) \approx \alpha \beta/\Gamma$.
Thus the relative velocity is negligible for $\Gamma \gg 1$
and points S$_\mathrm{P}$ and P are safely assumed to be in the same 
inertial frame.
The observed luminosity of TeV $\gamma$-rays from S is enhanced 
by a factor ${\cal D}_\xi^4$ because of the relativistic effects in Case 1,
where
\begin{eqnarray}
D_\xi &=& \frac{1}
{\Gamma [1-\beta \cos(\xi - \theta_\mathrm{obs})]} \\
&\approx& \frac{2}{1+(\alpha-\eta)^2} \, \Gamma
+ \frac{(\alpha-\eta)^4 
+ 6 (\alpha-\eta)^2 - 3}{6 [1+(\alpha-\eta)^2]^2} \,
\frac{1}{\Gamma}  + {\cal O}(\Gamma^{-3}) .
\end{eqnarray}
In Case 1, the ratio of the luminosity enhancements by the beaming effects
at S and P is 
$r = (D_\xi/D_\mathrm{obs})^4 
= \{(1+\eta^2)/[1+(\alpha-\eta)^2]\}^4$,
where
\begin{equation}
D_\mathrm{obs} = \frac{1}{\Gamma (1-\beta\cos\theta_\mathrm{obs})} 
\approx \frac{2}{1+\eta^2} \, \Gamma
+ \frac{\eta^4 + 6 \eta^2 - 3}{6 (1+\eta^2)^2} \,
\frac{1}{\Gamma} + {\cal O}(\Gamma^{-3}) .
\end{equation}
If $\alpha = 3$ and $\eta = 1$, 
$r \approx 1/40$.  In Case 2, $r = (D_\mathrm{obs}/D_\xi)^4 \approx 40$.


\section{Summary and Discussion}
\label{sec:sum}

In this paper we have proposed a structured jet model for the OTG flares.
The high energy emission spectra from TeV blazars have been well explained
by the leptonic jet models.
Because the leptonic jet models assume synchrotron self-Compton scattering
for the emission mechanism of very high energy $\gamma$-rays,
it is expected that flares occur in both X-ray and $\gamma$-ray energy
bands almost simultaneously.  
However, the OTG flares were not accompanied by
the simultaneous increase in the X-ray flux \citep{kra-et04,bla05},
and this is thought to be a challenging issue for the leptonic jet models. 
Our model of a structured jet assumes
that $\gamma$-rays are emitted in different components in the jet.
X-rays produced by a flare in a region, where
flares in X-rays and TeV $\gamma$-rays occur simultaneously,
are scattered in another region of the jet and the scattered photons are
observed as a secondary (OTG) flare.
Note that TeV $\gamma$-rays of the primary flare
are not scattered by electrons/positrons because of Klein-Nishina effects.
The emitted $\gamma$-ray energy is estimated as
$\epsilon'_\mathrm{SSC} = \gamma' m_e c^2$,
because the scattering occurs mainly in Klein-Nishina regime.
Then the observed $\gamma$-ray energy is roughly given by
\begin{equation}
\epsilon_\mathrm{SSC} \sim \Gamma \epsilon'_\mathrm{SSC}
\sim 5.5 \times 10^{12}
\left( \frac{B'}{0.1 \mathrm{G}} \right)^{-1/2}
\left( \frac{\Gamma}{20} \right)^{1/2}
\left( \frac{\epsilon_\mathrm{syn}}{10 \mathrm{keV}} \right)^{1/2}
 \quad \mathrm{eV} ,
\end{equation}
using equation (\ref{eq:gamma-value}).

\begin{figure}
\includegraphics[angle=0,scale=.55]{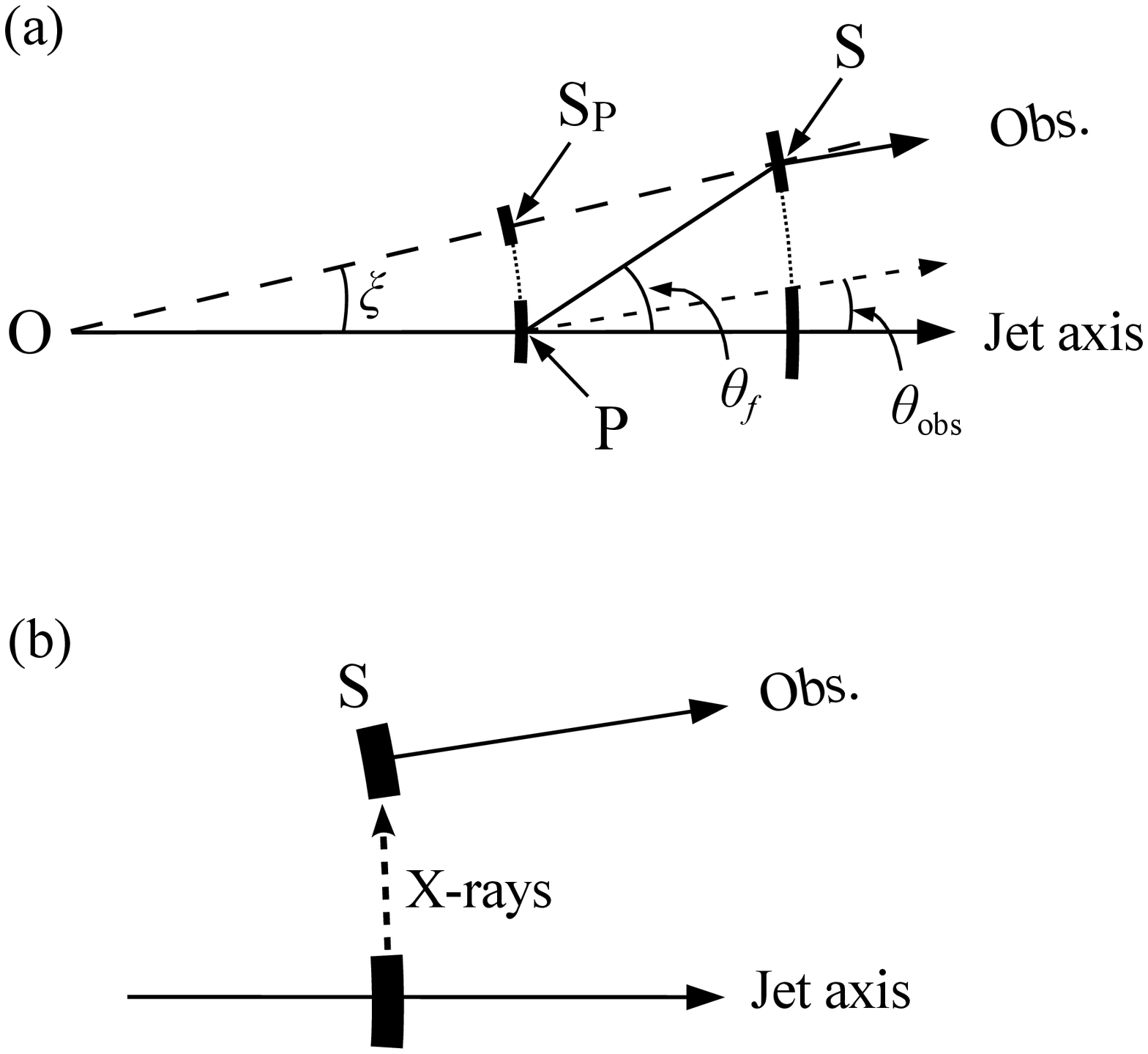}
\caption{Case 1.
(a) The jet is ejected from O and consists of patchy regions,
and they move radially with Lorentz factor $\Gamma$.
The primary flare occurs at P.  During time interval $\delta t$,
the dense region around S$_\mathrm{P}$ moves to S
and a fraction of X-rays emitted at P travels toward S.
Then the secondary (``orphan'') flare is triggered at S.
The observer is located at angle
$\theta_\mathrm{obs} = \eta/\Gamma$ off the jet axis,
and the points S$_\mathrm{P}$ and S are on the line
at angle $\xi = \alpha/\Gamma$ off the jet axis.
The distances OP and OS$_\mathrm{P}$ are $d$, 
$\mathrm{S}_\mathrm{P} \mathrm{S} = \beta c \delta t$,
and $\mathrm{PS} = c \delta t$.
(b) Emission regions in the comoving frame of the jet.
X-rays emitted in the primary flare are injected into S and they are
scattered by nonthermal particles in S.
\label{fig:jet-case1}}
\end{figure}

\begin{figure}
\includegraphics[angle=0,scale=.59]{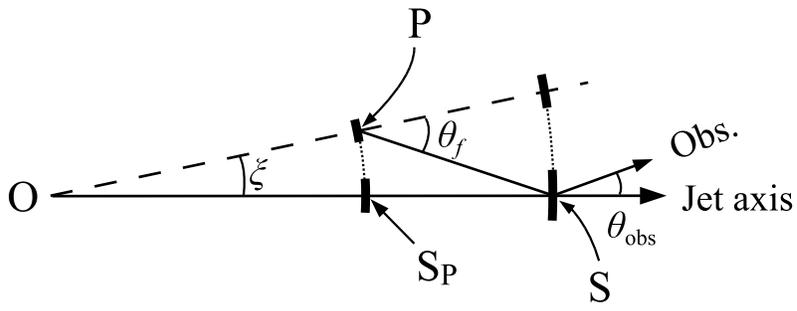}
\caption{Case 2. Same as Fig. \ref{fig:jet-case1} except 
that X-rays come from the off-axis region around P.
\label{fig:jet-case2}}
\end{figure}

Because the X-rays emitted in the primary flare need a propagation time 
before they are scattered in different parts of the jet,
this results in the time delay between the primary and secondary flares.
In our model, the delay time between the primary flare and 
the secondary TeV $\gamma$-ray flare is dependent on 
$\Gamma$, $\alpha$, $\eta$, and $d$.
The observed delay time is about 15 days for 1ES 1959+650
and this value is obtained for
$\Gamma = 20$, $\alpha = 3$, $\eta = 1$, and $d = 4 \times 10^{17}$ cm
in Case 1.
As the value of $\alpha$ decreases the value of $d$ increases for a given
value of $\delta t_\mathrm{obs}$.
If the value of $\Gamma$ is known from the observations of the primary flare, 
the values of $\alpha$ and $d$ might be reasonably guessed from the delay
time.
It is not clear from the numerical values estimated in this paper
whether Case 1 is favored to Case 2 or vice versa.
Rare observations of OTG flares might be
due to that this phenomenon needs a specific structure of the jets 
as shown in Figures \ref{fig:jet-case1} or \ref{fig:jet-case2}.

\begin{figure}
\includegraphics[angle=0,scale=.65]{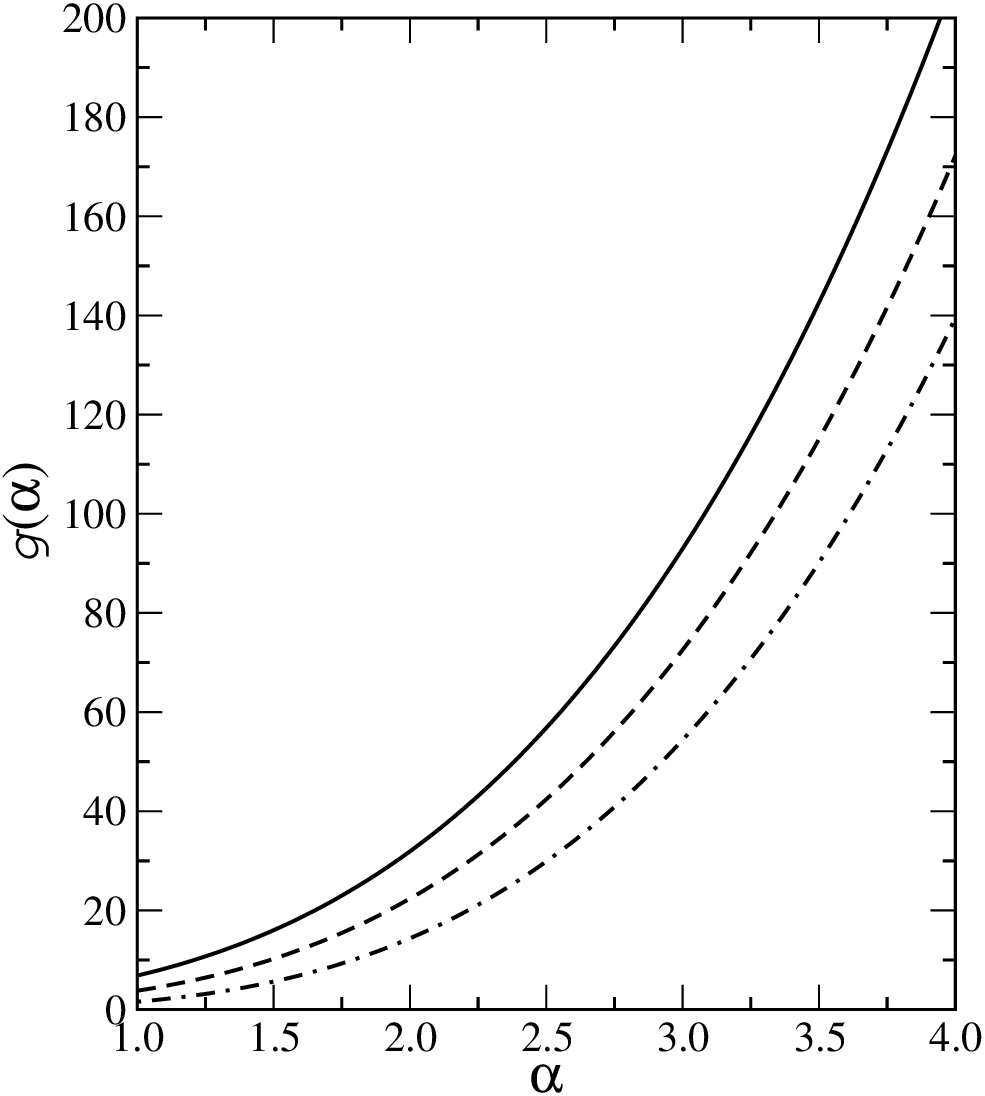}
\caption{$g(\alpha)$ for $\eta = 0$ (solid), 0.5 (dashed),
and 1 (dot-dashed).
\label{fig:g-factor}}
\end{figure}

The duration of the secondary flare depends on the time interval of
the primary flare and the angular size of the secondary flare site,
$\Delta \theta$.
For example, in Case 1, 
if $\Delta \theta$ is given by $\Delta \alpha /\Gamma$ and
the primary flare is impulsive,
the observed duration of the second flare is given by
the difference between the values of $\delta t_\mathrm{obs}$ for
$\alpha$ and $\alpha + \Delta \alpha$:
\begin{equation}
t_\mathrm{dur} 
\approx 0.29 \left( \frac{20}{\Gamma} \right)^2
\left( \frac{d}{3 \times 10^{17} \, \mathrm{cm} } \right) 
\frac{d f(\alpha)}{d \alpha}  \Delta \alpha \quad \mathrm{days} ,
\end{equation}
where $\Delta \alpha \ll 1$ is assumed.
The values of $g(\alpha) = df/d\alpha$ are shown in Figure \ref{fig:g-factor}
for Case 1.
For example, $g \approx 54.4$ for $\eta = 1$ and $\alpha = 3$.
If the observed duration of the secondary flare is about 5 hours,
$\Delta \alpha \sim 0.01$ for $\alpha = 3$, $\Gamma = 20$, 
and $d = 4 \times 10^{17}$ cm.
The observed duration of the secondary flare may depend on
not only the value of $\Delta \alpha$ but also the duration
of the primary flare, 
and the effect of the latter elongates the duration of the secondary flare.

We have shown that the time lag between the primary and secondary
flares are well accounted for by our model.
Next we argue that the density of soft photons (X-rays)
from the primary flare is high enough to produce the OTG flare
if the particle injection time in the primary flare is sufficiently long,
and that the particle injection rate needed in the secondary flare
is not unreasonably high.

First, it should be noted that photons from the primary flare blob (PB)
are not injected from the rear side of the secondary flare blob (SB).
In the comoving frame of the blobs, the angle between
the line of sight and the incident direction of photons from PB is
about 90 degrees.
Then the energy boost by inverse Compton scattering in SB should be effective,
although the detailed calculations of radiative transfer are necessary to
obtain the emission spectrum of TeV $\gamma$-rays from SB.

Next, we discuss the amount of radiation received by SB and the cooling time
of leptons.  In the following, primed symbols are quantities during 
the primary flare in the comoving frame and doubly primed symbols are 
quantities during the secondary flare in the comoving frame.
We assume that PB and SB are spherical plasma clouds for simplicity.
The radii of PB and SB in the comoving frame
at the occurrence of the primary flare are denoted by
$R'_\mathrm{prim}$ and $R'_\mathrm{sec}$, respectively.
The radius of SB at the time when the secondary flare occurs is denoted by
$R''_\mathrm{sec}$.
The solid angle extended by SB for a point in PB is given by
$\Omega'_\mathrm{sec} \sim \pi R^{\prime \, 2} _\mathrm{sec}
/d^{\prime \, 2}_\mathrm{sec}$,
where $d'_\mathrm{sec}$ is the distance between PB and SB:
$d'_\mathrm{sec} \sim \xi d = (\alpha/\Gamma) \, d$.
If SB expands proportionally with the distance from the central black hole,
$\Omega'_\mathrm{sec} = \Omega''_\mathrm{sec}$.
If radiation from PB is emitted isotropically
in the comoving frame, the fraction of the radiation received by SB
is given by
\begin{equation}
r_\mathrm{rad} \equiv 
\frac{\Omega'_\mathrm{sec}}{4 \pi} 
= \frac{1}{4} \, \frac{\Gamma^2}{\alpha^2} \,
\frac{R^{\prime \, 2}_\mathrm{sec}}{d^2}
\sim  \frac{1}{36}
\left(\frac{\Gamma}{20}\right)^2
\left(\frac{3}{\alpha}\right)^2
\left(\frac{R'_\mathrm{sec}}{2 \times 10^{16} \mathrm{cm}} \right)^2
\left(\frac{4 \times 10^{17} \mathrm{cm}}{d}\right)^2 .
\end{equation}
We assumed $R'_\mathrm{sec} \sim 2 \times 10^{16}$ cm;
this value is close to the values assumed for the emission region of the flare models 
of 1ES 1959+650 in \citet{kra-et04}.
Thus the soft photon (X-ray) energy density in SB is roughly
about 1/30 of that in PB and the Compton cooling time of nonthermal particles
in SB is longer by 30 times than in PB.
On the other hand, the flare duration in PB is determined by the
injection duration of nonthermal particles, because the cooling time
should be short owing to the large energy densities of soft photons and magnetic
fields.  Thus the observed OTG flare is explained,
if the particle injection duration in PB
is about 30 times longer than the cooling time in SB.
Furthermore, the energy densities of magnetic fields and internal synchrotron
photons in SB are smaller than in PB, so that emission other than TeV $\gamma$-rays
is too weak to be observed.

In the following we estimate the required injection rate
of nonthermal leptons in the secondary flare.
The observed energy flux of the secondary flare at 600GeV
is $\nu F_\nu \sim 3 \times 10^{-10}$ ergs s$^{-1}$ cm$^{-2}$.
The luminosity distance to 1ES 1959+650 ($z = 0.047$) is $d_L = 210$ Mpc,
assuming that $H_0 = 70$ km s$^{-1}$ Mpc$^{-1}$,
$\Omega_\Lambda = 0.7$, and $\Omega_m = 0.3$, where $H_0$,
$\Omega_\Lambda$, and $\Omega_m$ are the Hubble constant and
the density parameters of dark energy and matter, respectively.
Then the luminosity of the very high
energy gamma-rays is $L_\mathrm{VHE, sec} \sim 1.6 \times 10^{45}$ ergs s$^{-1}$.
The luminosity in the comoving frame is given by
\begin{equation}
L''_\mathrm{VHE, sec} = \frac{L_\mathrm{VHE, sec}}{{\cal D}^4_\mathrm{sec}}
\sim
9.9 \times 10^{39} \left( \frac{{\cal D}_\mathrm{sec}}{20} \right)^{-4} \quad
\mathrm{ergs} \, \mathrm{s}^{-1} ,
\end{equation}
where ${\cal D}_\mathrm{sec}$ is the beaming factor of SB
and ${\cal D}_\mathrm{sec} = 8$ and 20 for Cases 1 and 2, respectively,
if $\Gamma = 20$, $\alpha = 3$, and $\eta = 1$.
The distance between the central black hole and SB is
$d_\mathrm{sec} = d + \beta c \delta t \sim 11 d$,
where $\alpha = 3$ is assumed.
Assume that SB expands linearly with the distance from the 
central black hole.  Then the radius of SB is given by
\begin{equation}
R''_\mathrm{sec}  = \frac{d_\mathrm{sec}}{d} \, R'_\mathrm{sec} 
\sim 11 R'_\mathrm{sec}
= 2.2 \times 10^{17} \left( \frac{R'_\mathrm{sec}}
{2 \times 10^{16} \mathrm{cm}} \right) \quad \mathrm{cm} .
\end{equation}
We obtain the energy injection rate per unit volume 
during the secondary flare:
\begin{equation}
q''_\mathrm{sec} = \frac{L''_\mathrm{VHE, sec}}{\frac{4 \pi}{3} 
R^{\prime \prime \, 3}_\mathrm{sec} } 
= 2.2 \times 10^{-13} 
\left( \frac{R'_\mathrm{sec}}{2 \times 10^{16} \mathrm{cm}} \right)^{-3}
\left( \frac{{\cal D}_\mathrm{sec}}{20} \right)^{-4} 
\quad \mathrm{ergs} \, \mathrm{s}^{-1} \, \mathrm{cm}^{-3} .
\end{equation}
The injection rate of nonthermal leptons in SB is also given by
\begin{equation}
N''_e = \frac{L''_\mathrm{VHE, sec}}{\langle \gamma \rangle m_e c^2} 
= 2.4 \times 10^{40} \left( \frac{{\cal D}_\mathrm{sec}}{20} \right)^{-4} 
\left(\frac{\langle \gamma \rangle}{5 \times 10^5} \right)^{-1} \quad
\mathrm{s}^{-1} ,
\end{equation}
where $\langle \gamma \rangle$ is the value of the characteristic Lorentz
factor of injected leptons in SB.
When the energy injection rate during the primary flare is denoted by
$q'_\mathrm{prim}$,
we obtain
\begin{equation}
\frac{q''_\mathrm{sec}}{q'_\mathrm{prim}}
= \frac{L''_\mathrm{VHE, sec}
/\left(\frac{4 \pi}{3} R^{\prime \prime \, 3}_\mathrm{sec}\right)}
{L'_\mathrm{VHE, prim}
/\left(\frac{4 \pi}{3} R^{\prime \, 3}_\mathrm{prim}\right)}
= \frac{[\nu F_\nu]_\mathrm{VHE, sec}}{[\nu F_\nu]_\mathrm{VHE, prim}} 
\left(\frac{{\cal D}_\mathrm{prim}}{{\cal D}_\mathrm{sec}}\right)^4
\left(\frac{R'_\mathrm{prim}}{R''_\mathrm{sec}}\right)^3 ,
\end{equation}
where $[\nu F_\nu]_\mathrm{VHE, prim}$ and $[\nu F_\nu]_\mathrm{VHE, sec}$
are the observed quantities of $\nu F_\nu$ in the primary and secondary flares,
respectively.
If $R'_\mathrm{sec} \sim R'_\mathrm{prim}$ and 
$R''_\mathrm{sec} \sim 11 R'_\mathrm{sec}$ are assumed,
$q''_\mathrm{sec} / q'_\mathrm{prim}
\sim  c_1 \, [\nu F_\nu]_\mathrm{VHE, sec}
/ [\nu F_\nu]_\mathrm{VHE, prim}$,
where $c_1 \sim  3.0 \times 10^{-2}$ and
$1.9\times 10^{-5}$ for Cases 1 and 2, respectively.
From the observed data \citep{kra-et04},
we may set \\ \noindent
$[\nu F_\nu]_\mathrm{VHE, sec}/[\nu F_\nu]_\mathrm{VHE, prim} \sim 1$,
and $q''_\mathrm{sec} \sim c_1 q'_\mathrm{prim}$ is obtained.
Thus the required value of $q''_\mathrm{sec}$ is not unreasonably large.
This particle energy is transfered to X-rays by
inverse Compton scattering to emit TeV $\gamma$-rays in the secondary flare.

Our model assumes a structured jet and the occurrence
of OTG flares depends on the geometry of dense regions in the jet.
Alternatively the dense regions might be widely distributed,
e.g., the layer-spine model by \citet{gtc05},
and the value of $\alpha$ depends on where the region of particle
acceleration at pc-scale exists.

Finally we comment on the observed large scale structure of jets.
The line of sight is inside the opening angle of jets in our model,
and this may result in a core-halo structure of the emission region.
However, since the emission regions of high energy radiation are
located very close to the central black hole,
the large scale structure of jets is affected by the environment
of the jets.  That is, the interaction between jets and gases surrounding
the central region may cause the bending of the jets.
It is also possible that the opening angle may change with the distance from
the central region.  These may account for the jet structure
observed by VLBI.

\acknowledgments
We thank the anonymous referee for the valuable comments.
This work has been partially supported by Scientific Research Grants
(F. T.: 14079205 and 16540215) from
the Ministry of Education, Culture, Sports, Science and Technology of Japan.


\clearpage

\end{document}